\documentclass[12pt]{article}
  \usepackage{amsfonts}
  \usepackage{amsmath}
\usepackage{amssymb}
\usepackage{amscd}
\usepackage[dvips]{graphicx}
\begin{document}

~~
\bigskip
\bigskip
\begin{center}
{\Large {\bf{{{Kerr black hole in canonically deformed space-time}}}}}
\end{center}
\bigskip
\bigskip
\bigskip
\begin{center}
{{\large ${\rm {Marcin\;Daszkiewicz}}$}}\\
\bigskip
{ ${\rm{Institute\; of\; Theoretical\; Physics}}$}

{ ${\rm{ University\; of\; Wroclaw,\; pl.\; Maxa\; Borna\; 9,\;
50-206\; Wroclaw,\; Poland}}$}

{ ${\rm{ e-mail:\; marcin@ift.uni.wroc.pl}}$}
\end{center}
\begin{center}
{{\large ${\rm {Cezary\;J.\;Walczyk}}$}}\\
\bigskip
{ ${\rm{Department\; of\; Physics}}$}

{ ${\rm{ University\; of\; Bialystok,\; ul.\; Lipowa\;41,\; 15-424\; Bialystok,\; Poland}}$}

{ ${\rm{ e-mail:\; c.walczyk@alpha.uwb.edu.pl}}$}
\end{center}
\bigskip
\bigskip
\bigskip
\bigskip
\bigskip
\bigskip
\bigskip
\bigskip
\bigskip
\begin{abstract}
We investigate the Kerr black hole defined on canonically deformed space-time. Particulary, we find the corresponding event horizon,
the ergosphere, the temperature and the entropy of  such deformed object.
\end{abstract}
\bigskip
\bigskip
\bigskip
\bigskip
\eject

\section{{{Introduction}}}
The idea to use noncommutative coordinates is quite old - it goes
back to Heisenberg and was firstly formalized by Snyder in
\cite{snyder}. Recently, however,  there were  found new formal
arguments based mainly on Quantum Gravity \cite{2}, \cite{2a} and
String Theory models \cite{recent}, \cite{string1}, indicating that
space-time at Planck scale  should be noncommutative, i.e. it should
have a quantum nature. On the other side, the main reason for such
considerations follows from many phenomenological considerations,
which state that relativistic space-time symmetries should be
modified (deformed) at Planck scale, while  the classical Poincare
invariance still remains valid at larger distances
\cite{1a}, \cite{1d}.

 In accordance with the Hopf-algebraic classification of all
deformations of relativistic and nonrelativistic symmetries (see
\cite{clas1}, \cite{clas2}) the most general form of space-time
noncommutativity looks as follows
\begin{equation}
[\,x_{\mu},x_\nu\,] = \theta_{\mu\nu} (x)\;,\label{nonco}
\end{equation}
 where
 \begin{equation}
 \theta_{\mu\nu} (x) = \theta_{\mu\nu}^{(0)} + \theta_{\mu\nu}^{(1)\,\rho}x_{\rho} +
 \theta_{\mu\nu}^{(2)\,\rho\tau}x_{\rho}x_{\tau}\;.\label{par}
\end{equation}

For the simplest, canonical noncommutativity $(\theta_{\mu\nu} (x) =
\theta_{\mu\nu}^{(0)})$, the corresponding Poincare Hopf algebra has
been provided in \cite{oeckl} and \cite{chi} with the use of  twist
procedure \cite{drin}-\cite{twist1}, while its nonrelativistic
counterparts have been discovered by various contraction schemes in
\cite{dasz}.

The Lie-algebraic $(\theta_{\mu\nu} (x)
=\theta_{\mu\nu}^{(1)\,\rho}x_{\rho})$ relativistic and
nonrelativistic symmetries have been proposed in \cite{kappaP} and
\cite{kappaG} respectively. In the literature they are known as
$\kappa$-Poincare and $\kappa$-Galilei Hopf algebra
with   mass-like deformation parameter $\kappa$.
Besides, there were proposed the twist deformations of a Lie-type at
relativistic and nonrelativistic level in \cite{lie2}, \cite{lie1}
and \cite{dasz}.

The quadratic deformation $(\theta_{\mu\nu} (x)
=\theta_{\mu\nu}^{(2)\,\rho\tau}x_{\rho}x_{\tau})$ has been studied
in \cite{paolo} and \cite{lie2}.

Recently, there appeared a lot of papers dealing with classical
(\cite{deri}-\cite{daszwal}) and quantum (\cite{qm1}-\cite{oscy})
mechanics, Doubly Special Relativity frameworks (\cite{dsr1a},
\cite{dsr1b}), statistical physics (\cite{maggiore}, \cite{rama})
and field theoretical models (see e.g. \cite{przeglad}), defined on
the canonically and Lie-algebraically deformed
 space-times\footnote{For earlier studies see \cite{lukiluk1} and
\cite{lukiluk2}.}. It should be noted, however, that the especially interesting studies have been performed in  articles \cite{treatment}-\cite{inne2}   in the context
of so-called black hole physics, i.e.
there have been investigated (with use of different methods and techniques) the basic types of noncommutative black hole metrics.

In this article we study the effect of canonical noncommutativity  on the Kerr solution of General Relativity equation with use of  treatment proposed in \cite{treatment}-\cite{treatment2}. Particulary, we investigate the impact of quantum space on the event horizon as well as on the shape of
ergosphere. Besides, as in the case of  basic nonrotating black holes described by Schwarzschild and Reisner-Nordstrom metric tensors
 respectively,  we find the temperature and
entropy  of  such canonically  deformed object.

 In our investigation we proceed in accordance with the following algorithm (see \cite{treatment}-\cite{treatment2})\footnote{It should be noted, however, that the
 results obtained by us make full solutions in deformation parameter $\theta$ of singularity equations, contrary to the results of articles \cite{treatment}-\cite{treatment2}.}.  Firstly, we assume that particle-like gravitational source remains point-like. Next, we exchange the commutative variables in classical Kerr solution of field equation by noncommutative ones. Further, we rewrite the obtained in such a way metric tensor in terms of commutative phase space variables and deformation parameter $\theta$ (see formula (\ref{rep})). Finally, we perform the basic analysis of    such a prepared mathematical  object.

It should be  noted that the classically noncommutative Kerr black hole has been  already analyzed   in article \cite{inne2}. However, the used therein
techniques are completely different than the methods preferred by us. For example, there  is  assumed  that due to the space-time noncommutativity
the mass density  particle-like gravitational source is smeared. Such an assumption leads to the new (Kerr) solution of General Relativity equations for which
the corresponding event horizon can be found only numerically.

The paper is organized as follows. In Sect. 2 we recall basic facts
concerning the canonical deformed Poincare Hopf algebra and the corresponding quantum   space-time
provided in article \cite{chi}. The third section is devoted to the short review of Kerr black hole defined in
commutative (classical) space. In Sect. 4 we analyze the effect of space-time  noncommutativity (\ref{minkowski}) on the
Kerr metric.  The final remarks are presented in the last section.

\section{Canonically deformed Minkowski space-time}

In this section we recall  basic facts associated with $\theta$-deformed Poincare Hopf algebra ${\cal U}_{\theta}({P})$
and with the corresponding canonically deformed quantum space-time \cite{chi}. Firstly, it should be noted that such  objects can be
get by so-called twist procedure, in which the algebraic sector of Hopf structure ${\cal U}_{\theta}({P})$ remains undeformed, i.e. it takes
the form
\begin{eqnarray}
&&\left[\; M_{\mu \nu },M_{\rho \sigma }\;\right] =i\left( \eta _{\mu
\sigma }\,M_{\nu \rho }-\eta _{\nu \sigma }\,M_{\mu \rho }+\eta
_{\nu \rho }M_{\mu
\sigma }-\eta _{\mu \rho }M_{\nu \sigma }\right) \;,  \notag \\
&&\left[\; M_{\mu \nu },P_{\rho }\;\right] =i\left( \eta _{\nu \rho
}\,P_{\mu }-\eta _{\mu \rho }\,P_{\nu }\right) \;\;\;,\;\;\; \left[\;
P_{\mu },P_{\nu }\;\right] =0\;.  \label{poincare}
\end{eqnarray}
Besides, the coproduct of considered algebra is given by
\begin{equation}
\Delta _{0}(a) \to \Delta _{\theta}(a) = \mathcal{F}_{\theta }\circ
\,\Delta _{0}(a)\,\circ \mathcal{F}_{\theta
}^{-1}\;,\label{twist}
\end{equation}
where
\begin{equation}
{\cal F}_{\theta}=\exp \left[\;{\frac{{ i}}{2}\theta^{\mu\nu}P_\mu\otimes P_\nu}\;\right]\;,\label{factor}
\end{equation}
denotes the canonical twist factor, while  $\Delta _{0}(a) = a \otimes 1 + 1 \otimes a$. Consequently, using (\ref{poincare})-(\ref{factor})
we get
\begin{eqnarray}
\Delta_{\theta}(P_\mu) &=& P_\mu\otimes 1\ +\ 1\otimes P_\mu\;, \nonumber \\
\Delta_\theta (M_{\mu\nu}) &=& M_{\mu\nu}\otimes 1\ +\ 1\otimes M_{\mu\nu}
 - \frac{1}{2}({\theta_\mu}^\rho P_\nu-{\theta_\nu}^\rho
P_\mu)\otimes P_\rho\ + \label{cop}\\
&+&\frac{1}{2}P_\rho\otimes({\theta_\mu}^\rho P_\nu-{\theta_\nu}^\rho
P_\mu)\;. \nonumber
\end{eqnarray}
The corresponding quantum Minkowski space-time is defined as the representation space (Hopf module) for Poincare Hopf algebra
${\cal U}_{\theta}({P})$. It is given by the following commutation relationes
\begin{equation}
[\;{\hat x}_\mu,{\hat x}_\nu\;] =\ { i}\theta_{\mu\nu}\;,\label{minkowski}
\end{equation}
and it can be extended to the whole algebra of momentum and position operators as follows
\begin{equation}
[\;{\hat x}_\mu,{\hat x}_\nu\;] =\ { i}\theta_{\mu\nu}\;\;\;,\;\;\; [\;{\hat p}_\mu,{\hat p}_\nu\;] =0\;\;\;,\;\;\;
[\;{\hat x}_\mu,{\hat p}_\nu\;]={ i}\eta_{\mu\nu}\;. \label{phase}
\end{equation}
Of course, for deformation parameter $\theta$ running to zero the all above objects become classical.

\section{{Kerr black hole}}

The Kerr metric describes the solution of  field equation for rotating (with angular momentum $L$), uncharged and axially-symmetric massive (with mass $M$) object in empty
space-time. In spherical coordinate system $(r,\varphi,\phi)$ it can be written as follows \cite{schw}
\begin{eqnarray}
c^{2} d\tau^{2} &=& \left( 1 - \frac{r_{s} r}{\rho^{2}} \right) c^{2} dt^{2} - \frac{\rho^{2}}{\Delta} dr^{2} - \rho^{2} d\varphi^{2}  + \label{kerr}\\
&-& \left( r^{2} + \alpha^{2} + \frac{r_{s} r \alpha^{2}}{\rho^{2}} \sin^{2} \varphi \right) \sin^{2} \varphi \ d\phi^{2}+\frac{2r_{s} r\alpha \sin^{2} \varphi }{\rho^{2}} \, c \, dt \, d\phi\;,\nonumber
\end{eqnarray}
with $r_s$ denoting the Schwarzschild radius
\begin{equation}
r_{s} = \frac{2GM}{c^{2}}\;,\label{rsradius}
\end{equation}
and with the length-scales $\alpha$, $\rho$ and $\Delta$ introduced for brevity
\begin{eqnarray}
\alpha &=& \frac{L}{Mc}\;,\label{alfa}\\
\rho^{2} &=& r^{2} + \alpha^{2} \cos^{2} \varphi\;,\label{ro}\\
\Delta &=& r^{2} - r_{s} r + \alpha^{2}\;.\label{delta}
\end{eqnarray}
Of course, for parameter $\alpha$ approaching zero  the above metric passes into well-known solution for Schwarzschild black hole
\cite{schw}. It should be  noted that the Kerr metric (\ref{kerr})  has four (only two of them are physically relevant) surfaces
on which it appears to be singular. First pair occurs when the purely radial component of metric goes to infinity. Then the solutions of
corresponding quadratic equation
\begin{equation}
\frac{1}{g_{rr}} = \frac{\Delta}{\rho^2} = 0\;,\label{equation1}
\end{equation}
look as follows
\begin{eqnarray}
r_{\rm inner} &:=& r_{i+} = \frac{r_{s} + \sqrt{r_{s}^{2} - 4\alpha^{2}}}{2}\;,\label{sol1}
\end{eqnarray}
and
\begin{eqnarray}
r_{i-} &=& \frac{r_{s} - \sqrt{r_{s}^{2} - 4\alpha^{2}}}{2}\;,\label{sol2}
\end{eqnarray}
respectively. The first of them (physical one) describes the event horizon while the second solution (due to the fact that radius $r_{i+}$ is bigger than $r_{i-}$) has just unphysical properties. \\
The second pair of singularities occurs when the purely temporal component of the Kerr metric changes sign. Again, solving a quadratic equation of the form
\begin{equation}
{g_{tt}} = \left( 1 - \frac{r_{s} r}{\rho^{2}} \right) = 0\;,\label{equation2}
\end{equation}
we get
\begin{eqnarray}
r_{\rm outer} &:=& r_{o+} =  \frac{r_{s} + \sqrt{r_{s}^{2} - 4\alpha^{2}\cos^2\varphi}}{2}\;,\label{sol3}\\
r_{o-} &=& \frac{r_{s} - \sqrt{r_{s}^{2} - 4\alpha^{2}\cos^2\theta}}{2}\;,\label{sol4}
\end{eqnarray}
where only radius $r_{o+}$ remains physical $(r_{o+}>r_{i+}>r_{o-})$. \\
One can observe that due to the $\cos^2 \varphi$ term in (\ref{sol3}) the outer sphere touches the inner one (see (\ref{sol1})) at the poles
of rotation axis, where $\varphi$ equals 0 or $\pi$; the space between these two surfaces is called the ergosphere. Obviously, for
parameter $\alpha$ running to zero  we have
 \begin{equation}
\lim_{\alpha\to 0}r_{\rm inner} = \lim_{\alpha\to 0}r_{\rm outer} = r_s\;,\label{limit}
\end{equation}
and the ergosphere disappears. \\
Finally, it should be mentioned that in accordance with article \cite{bh1}, \cite{bh2} the temperature and entropy of Kerr black hole are given by
\begin{equation}
T = \frac{\hbar}{4\pi kc}\frac{\left(r_{i+}-r_{i-}\right)}{\left(r_{i+}^2+\alpha^2\right)}\;,\label{temp}
\end{equation}
and
\begin{equation}
S = \frac{kc}{4\hbar}{\left(r_{i+}^2+\alpha^2\right)}\;,\label{ent}
\end{equation}
respectively, with symbol $k$ denoting the Boltzman's constant.

\section{Noncommutative Kerr black hole}

Following the treatment proposed in papers \cite{treatment}-\cite{treatment2} we define the  metric for Kerr black hole in $\theta$-deformed
 space-time with $\theta_{0i} =0$ by
\begin{eqnarray}
c^{2} d\tau^{2} &=& \left( 1 - \frac{r_{s} \sqrt{\hat{r}\hat{r}}}{\hat{\rho}\hat{\rho}} \right) c^{2} dt^{2} - \frac{\hat{\rho}\hat{\rho}}{\hat{\Delta}} d\hat{r}d\hat{r} - \hat{\rho}\hat{\rho} d\varphi^{2}  +\label{nonkerr}\\
 &-&\left( \hat{r}\hat{r} + {\alpha}^2 + \frac{r_{s} \sqrt{\hat{r}\hat{r}} {\alpha}^2}{\hat{\rho}\hat{\rho}} \sin^{2} \varphi \right) \sin^{2} \varphi \ d\phi^{2} +\frac{2r_{s} \sqrt{\hat{r}\hat{r}}{\alpha} \sin^{2} \varphi }{\hat{\rho}\hat{\rho}} \, c \, dt \, d\phi\;, \nonumber
\end{eqnarray}
where
\begin{eqnarray}
\hat{\rho}\hat{\rho} &=& \hat{r}\hat{r} + {\alpha}^2 \cos^{2} \varphi\;,\label{nonro}\\
\hat{\Delta} &=& \hat{r}\hat{r} - r_{s} \sqrt{\hat{r}\hat{r}} + {\alpha}^2\;,\label{nondelta}
\end{eqnarray}
and where the components of $\hat{r} = \sqrt {\hat{x}_1^2 + \hat{x}_2^2 + \hat{x}_3^2}$  satisfy the commutation relations (\ref{minkowski}). The solutions of
\begin{eqnarray}
\frac{\hat{\Delta}}{\hat{\rho}\hat{\rho}} &=&
\frac{\hat{r}\hat{r} - r_{s} \sqrt{\hat{r}\hat{r}} + {\alpha}^2}{\hat{r}\hat{r} + {\alpha}^2 \cos^{2} \varphi} = 0\;,
\label{nonequation1}\\
\left( 1 - \frac{r_{s} \sqrt{\hat{r}\hat{r}}}{\hat{\rho}\hat{\rho}} \right) &=&
\left( 1 - \frac{r_{s} \sqrt{\hat{r}\hat{r}}}{\hat{r}\hat{r} + {\alpha}^2 \cos^{2} \varphi} \right) = 0\;,\label{nonequation2}
\end{eqnarray}
are the singularities of the metric (\ref{nonkerr}). \\
In order to analyze the above system we represent the
noncommutative varibles $({\hat x}_i, {\hat p}_i)$ in terms of classical phase space  $({ x}_i, { p}_i)$ as  (see e.g.
\cite{kijanka}, \cite{lukiluk2})
\begin{equation}
{\hat x}_{i} = { x}_{i} - \frac{1}{2}\theta_{ij}p_j
\;\;\;,\;\;\; {\hat p}_{i}=
p_i\;, \label{rep}
\end{equation}
where
\begin{equation}
[\;x_i,x_j\;] = 0 =[\;p_i,p_j\;]\;\;\;,\;\;\; [\;x_i,p_j\;]
={i}\delta_{ij}\;. \label{classpoisson}
\end{equation}
Then, the equations (\ref{nonequation1}) and (\ref{nonequation2}) take the form
\begin{equation}
\frac{\hat{\Delta}}{\hat{\rho}\hat{\rho}} = \frac{\left(x_i - \frac{1}{2}\theta_{ij}p_j\right)\left(x_i - \frac{1}{2}\theta_{ik}p_k\right)-r_s\sqrt{\left(x_i - \frac{1}{2}\theta_{ij}p_j\right)\left(x_i - \frac{1}{2}\theta_{ik}p_k\right)}+\alpha^2}
{\left(x_i - \frac{1}{2}\theta_{ij}p_j\right)\left(x_i - \frac{1}{2}\theta_{ik}p_k\right) +\alpha^2\cos^2 \varphi} =0\;, \label{classnonequation1}
\end{equation}
and
\begin{equation}
\left( 1 - \frac{r_{s} \sqrt{\hat{r}\hat{r}}}{\hat{\rho}\hat{\rho}} \right) =
\left( 1 - \frac{r_{s} \sqrt{\left(x_i - \frac{1}{2}\theta_{ij}p_j\right)\left(x_i - \frac{1}{2}\theta_{ik}p_k\right)}}{\left(x_i - \frac{1}{2}\theta_{ij}p_j\right)\left(x_i - \frac{1}{2}\theta_{ik}p_k\right) +\alpha^2\cos^2 \varphi} \right) =0
\;, \label{classnonequation2}
\end{equation}
respectively. First of all, one should notice that for $\alpha$ running to zero the above conditions become the same, and we reproduce the
noncommutative Schwarzschild black hole singularity equation given by \cite{treatment}, \cite{treatment1}
\begin{equation}
\frac{\hat{\Delta}}{\hat{\rho}\hat{\rho}} = \left( 1 - \frac{r_{s} \sqrt{\hat{r}\hat{r}}}{\hat{\rho}\hat{\rho}} \right) =
 1 - \frac{r_{s}}{ \sqrt{\hat{r}\hat{r}}}  = 0\;. \label{toschwarz}
\end{equation}
Further,  we rewrite the formulas  (\ref{classnonequation1}) and (\ref{classnonequation2}) as follows
\begin{eqnarray}
\frac{\hat{\Delta}}{\hat{\rho}\hat{\rho}} &=& \frac{\tilde{r}^2-r_s \tilde{r}+\alpha^2}{\tilde{r}^2+\alpha^2\cos^2\varphi}=0\;, \label{taylor1}\\
\left( 1 - \frac{r_{s} \sqrt{\hat{r}\hat{r}}}{\hat{\rho}\hat{\rho}} \right) &=&
 1 - \frac{r_s\tilde{r}}{\tilde{r}^2+\alpha^2\cos^2\varphi}=0 \;,\label{taylor2}
\end{eqnarray}
where
\begin{equation}
\tilde{r}=\sqrt{r^2-\frac{1}{2}\vec{\theta}\vec{L}+\frac{1}{16}\left(\vec{p}\times\vec{\theta}\right)^2}\;\;\;,\;\;\;\vec{L}=\vec{x}\times\vec{p} \;\;\;\;\;{\rm and}\;\;\;\;\;\theta_{ij}=\frac{1}{2}\epsilon_{ijk} \theta_k\;,\label{rTilde}
\end{equation}
as well as we find  the corresponding  solutions in the form
\begin{equation}
\tilde{r}_\pm(a) = \frac{r_s\pm\sqrt{r_s^2-4a}}{2}\;,
\end{equation}
with  $a=\alpha^2$ and $a=\alpha^2\cos^2\varphi$ in the case of equations (\ref{taylor1}) and  (\ref{taylor2}) respectively.\\
Hence, in accordance with the formula (\ref{rTilde}) we  get
\begin{equation}
r_\pm(a)=\frac{1}{2}\sqrt{\left(r_s\pm\sqrt{r_s^2-4 a}\right)^2+2\vec{\theta}\vec{L}-\frac{1}{4}\left(\vec{p}\times\vec{\theta}\right)^2}\;,
\label{rTaylor12}
\end{equation}
while in the limit of commutative space, one obtains
\begin{eqnarray}
r_{-}(\alpha^2) &=& r_{i-}\;,\label{limits2}\\
r_{+}(\alpha^2) &=& r_{i+} \;=\; r_{\rm inner}\;,\label{limits3}\\
r_{-}(\alpha^2\cos^2\varphi) &=& r_{o-}\;,\label{limits5}\\
r_{+}(\alpha^2\cos^2\varphi) &=& r_{o+} \;=\;  r_{\rm outer}\;.\label{limits6}
\end{eqnarray}
Consequently, due to the above limits we define the $\theta$-deformed ergosphere as the space occurring between radiuses $r_{+}(\alpha^2)$ and $r_{+}(\alpha^2\cos^2\varphi)$, i.e.
as the region existing between deformed outer radius and "noncommutative" event horizon. Besides, one can observe that for $\alpha$ and $\theta_{ij}$ approaching zero, we reproduce the Schwarzschild event horizon
\begin{equation}
\lim_{\alpha, \theta \to 0}r_{+}(\alpha^2) = \lim_{\alpha, \theta \to 0}r_{+}(\alpha^2\cos^2\varphi) = r_s \;,
\end{equation}
with remaining  radiuses vanishing. \\
Finally, it should be mentioned that in accordance with formulas (\ref{temp}) and (\ref{ent}) as well as due to the limits (\ref{limits2})-(\ref{limits6}), the temperature and entropy of noncommutative
Kerr black hole are given by
\begin{eqnarray}
T = \frac{\hbar}{4\pi kc}\frac{\left({r}_{+}(\alpha^2)-r_{-}(\alpha^2)\right)}{\left(r_{+}^2(\alpha^2)+\alpha^2\right)}\;,\label{ntemp}
\end{eqnarray}
and
\begin{eqnarray}
S = \frac{kc}{4\hbar}{\left(r_{+}^2(\alpha^2)+\alpha^2\right)}\;,\label{nent}
\end{eqnarray}
respectively.

\section{Final remarks}

In this article we investigate the Kerr black hole defined on canonically deformed space-time. Particulary, we find the corresponding event
horizon, the proper ergosphere, the temperature and the entropy of such  deformed object. Besides,  for parameters  $\alpha$ and $\theta_{ij}$
approaching zero we reproduce the classical  Schwarzschild black hole solution, while in the limit of commutative space
we arrive to the case of undeformed Kerr black hole metric tensor; the  presented
studies has been performed with use of methods and techniques proposed in articles \cite{treatment}, \cite{treatment1} and \cite{treatment2}.

\section*{Acknowledgments}
The authors would like to thank M. Szczachor, P. Gusin and J. Lukierski
for valuable discussions. This paper has been financially  supported  by Polish
NCN grant No 2011/01/B/ST2/03354.

\end{document}